\documentclass[runningheads,a4paper]{llncs}

\newcommand{\ag}{\gamma}

\usepackage{amssymb}
\usepackage{multirow}
\usepackage{graphicx}

\usepackage{url}
\newcommand{\keywords}[1]{\par\addvspace\baselineskip
\noindent\keywordname\enspace\ignorespaces#1}

\begin{document}

\mainmatter  % start of an individual contribution

\title{Conflict Solution According to ``Aggressiveness'' of Agents in Floor-Field-Based Model}
\titlerunning{``Aggressiveness'' in Floor-Field Model}
\author{Pavel Hrab\'ak\inst{1}\fnmsep\thanks{Corresponding author} \and Marek Buk\'a\v cek\inst{2}}
\authorrunning{P. Hrab\'ak M. Buk\'a\v cek}
\institute{Institute of Information Theory and Automation\\Czech Academy of Sciences\\Pod Vodarenskou vezi 4, 182 08 Prague, Czech Republic\\
\email{hrabak@utia.cas.cz}
\and
Faculty of Nuclear Sciences and Physical Engineering\\
Czech Technical University in Prague\\
Trojanova 13, 120 00 Prague, Czech Republic\\
\email{bukacma2@fjfi.cvut.cz}}

\maketitle
\begin{center}{\today}\end{center}

\begin{abstract}
This contribution introduces an element of ``aggressiveness'' into the Floor-Field based model with adaptive time-span. The aggressiveness is understood as an ability to win conflicts and push through the crowd. From experiments it is observed that this ability is not directly correlated with the desired velocity in the free flow regime. The influence of the aggressiveness is studied by means of the dependence of the travel time on the occupancy of a room. A simulation study shows that the conflict solution based on the aggressiveness parameter can mimic the observations from the experiment.
\keywords{Floor-Field model, conflict solution, aggressiveness.}
\end{abstract}

\section{Introduction}

This article focuses on a microscopic study of a simulation tool for pedestrian flow. The object of the study is a simulation of one rather small room with one exit and one multiple entrance, which may be considered as one segment of a large network. The behaviour of pedestrians in such environment has been studied by our group by means of variety experiments~\cite{BukHraKrb2015TGF,BukHraKrb2014Procedia} from the view of the boundary induced phase transition (this has been studied theoretically for Floor-Field model in~\cite{EzaYanNis2013JCA}). Observing data from these experiments we have found out that each participant has different ability to push through the crowd. Therefore, this article is motivated by the aim to mimic such behaviour by simple cellular model, which may be applied in simulations of apparently heterogeneous scenarios as~\cite{KleWas2014LNCS,SpaGeoSir2014LNCS}.

The original model is based on the Floor-Field Model~\cite{BurKlaSchZitPhysicaA2001,KirSch2002PhysicaA,Kretz2007PhD} with \emph{adaptive time-span}~\cite{BukHraKrb2014LNCS} and \emph{principle of bonds}~\cite{HraBukKrb2013JCA}. The adaptive time span enables to model heterogeneous stepping velocity of pedestrians; the principle of bonds helps to mimic collective behaviour of pedestrians in lines. It is worth noting that there is a variety of modifications of the Floor-Field model capturing different aspects of pedestrian flow and evacuation dynamics. Quite comprehensive summary can be found in~\cite{SchChoNis2010}.

In this article we focus on the solution of conflicts, which accompany all cellular models with parallel update, i.e., when more agents decide to enter the same site/cell. In such case, one of the agents can be chosen at random to win the conflict, the randomness can be executed proportionally to the hopping probability of conflicting agents~\cite{BurKlaSchZitPhysicaA2001}.  The unresolved conflicts play an important role in models of pedestrian evacuation. The aim to attempt the same cell may lead to the blocking of the motion. This is captured by the friction parameter $\mu$ denoting the probability that none of the agent wins the conflict. An improvement is given by the friction function~\cite{YanKimTomNisSumOhtNish2009PRE}, which raises the friction according to the number of conflicting agents.

In our approach we introduce an additional property determining the agent's ability to win conflicts, which may be understood as agent's aggressiveness. This characteristics has been inspired by the analyses of repeated passings of pedestrians through a room under various conditions from free flow to high congestion. As will be shown below, this characteristics significantly affects the time spent by individual agents in the room, which is referred to as the \emph{travel time}. Similar heterogeneity in agents behaviour has been used in~\cite{JiZhouRan2013PhysicaA}, where the ``aggressiveness'' has been represented by the willingness to overtake.

\section{Experiment}

\begin{figure}[h!]
	\begin{center}
	\hfill\includegraphics[height=.27\textwidth]{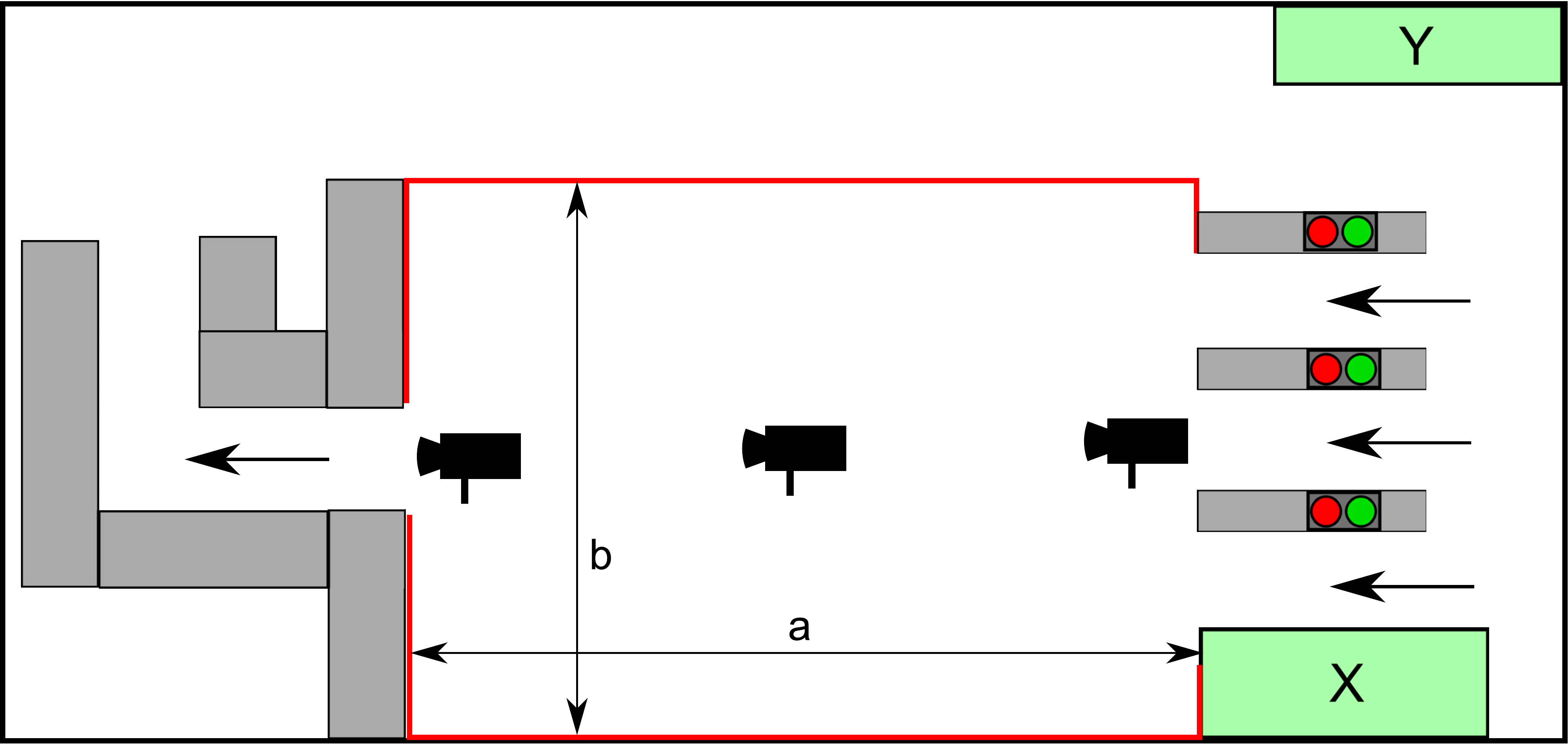}\hfill
	\includegraphics[height=.27\textwidth]{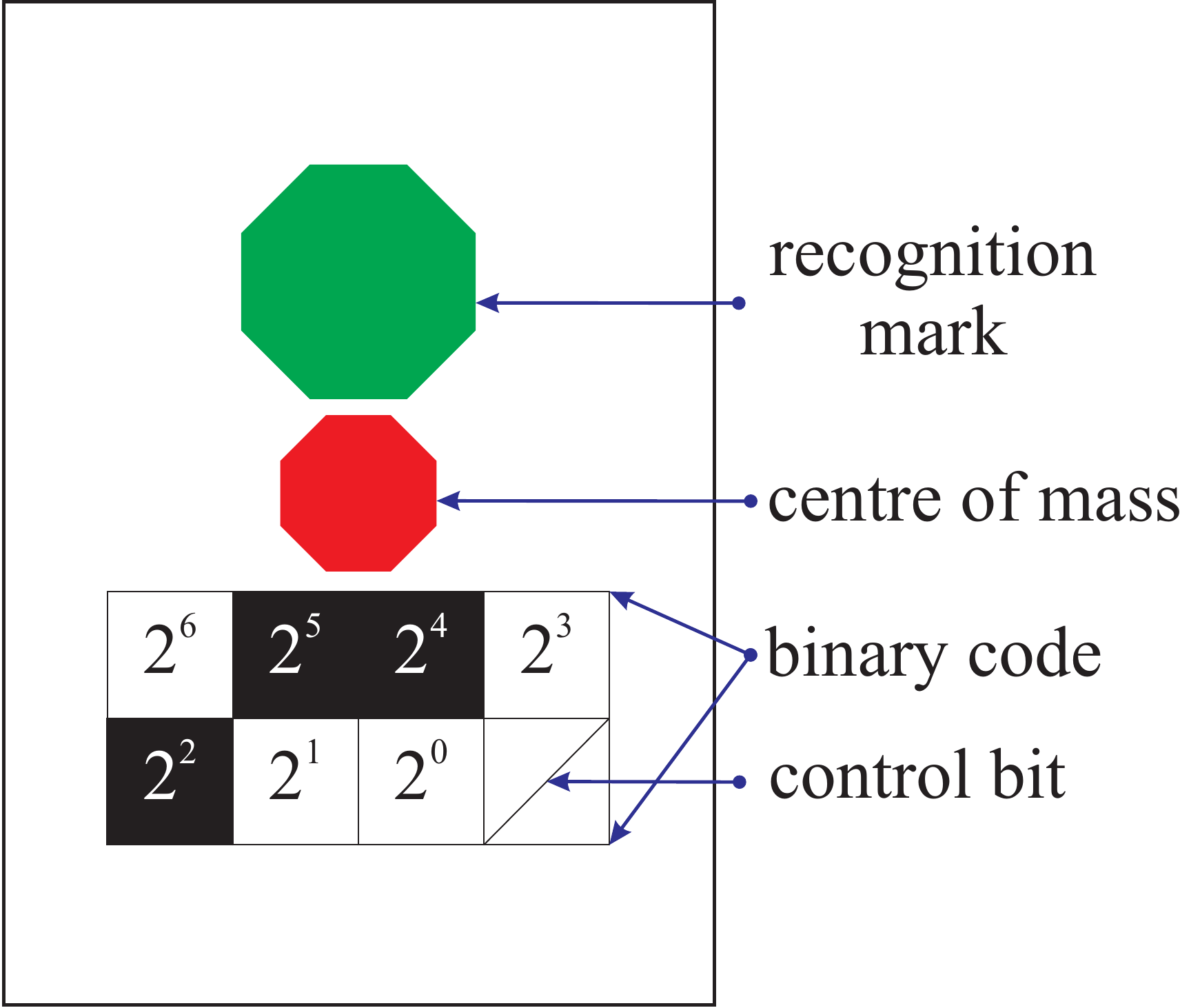}\hfill\phantom{x}
	\end{center}
\caption{Taken from~\cite{BukHraKrb2014Procedia}. Left: Experimental setting the experiment, a = 7.2~m, b = 4.4~m. After the passage through the exit, participant returned to the area Y waiting for another entry. Right: sketch of pedestrian's hat used for automatic image recognition.}
\label{fig:setting}
\end{figure}

The introduction of the aggressiveness as an additional model parameter is motivated by the microscopic analyses of the experimental data from the experiment ``passing-through'' introduced in~\cite{BukHraKrb2014Procedia}. The set-up of the experiment is shown in Figure~\ref{fig:setting}. Participants of the experiment were entering a rectangle room in order to pass through and leave the room via the exit placed at the opposite wall to the entrance. 

The inflow rate of pedestrians has been controlled in order to study the dependence of the phase (free flow or congested regime) on the inflow rate $\alpha$. In order to keep stable flow through the room, pedestrians were passing the room repeatedly during all runs of the experiment.

Each participant has been equipped by a hat with unique binary code. The automatic image recognition enables us not only to restore the pedestrians trajectories but more over to assign all trajectories to individual participants. This enables the study of individual properties of the pedestrians under a bride scale of conditions, since for each participant there are 20 to 40 records of their passings.

\begin{figure}
\centering
	\includegraphics[scale=.8]{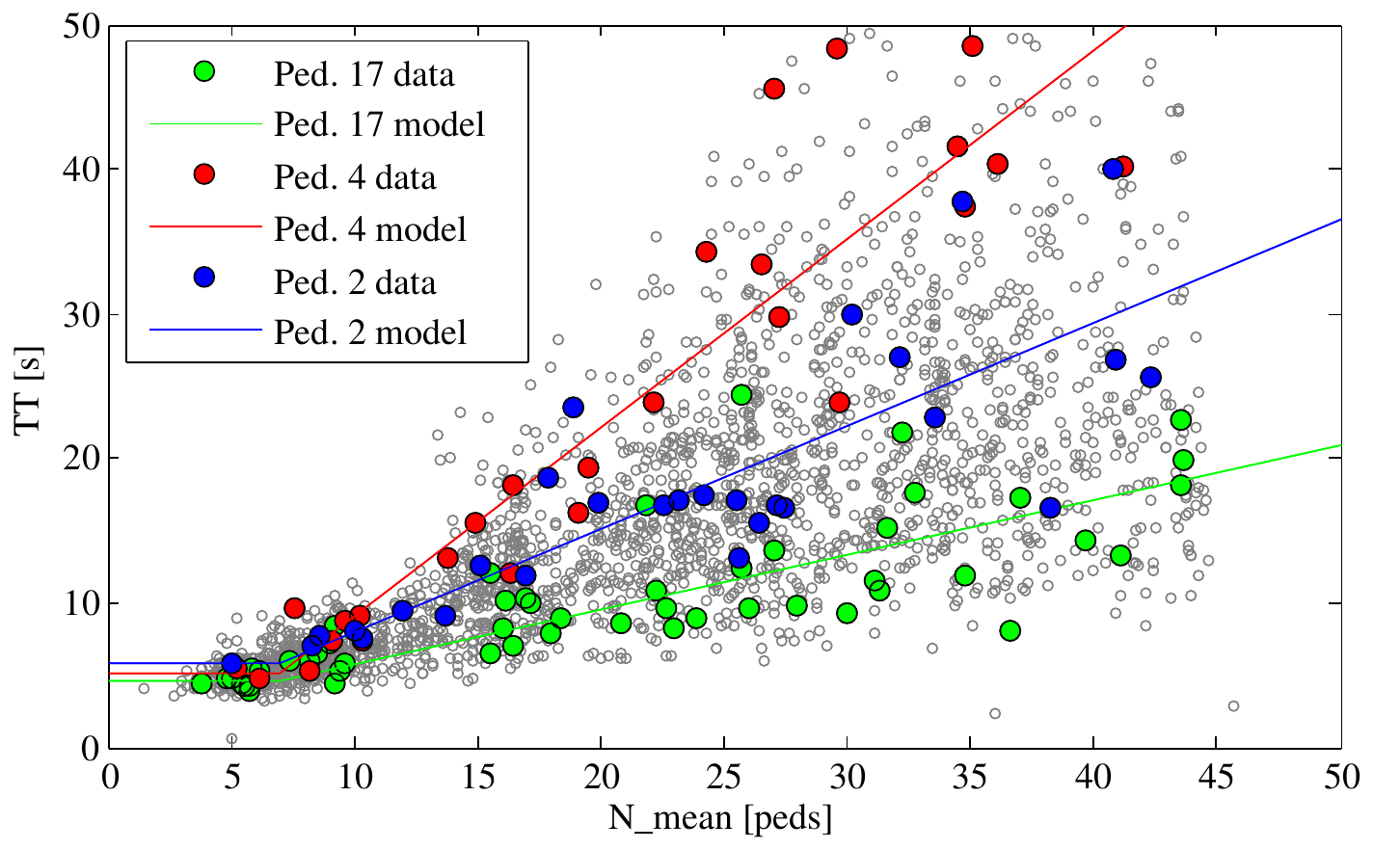}
\caption{Scatter plot of the travel time $TT$ with respect to the occupancy $N_\mathrm{mean}$ extracted from the experiment. Three participants are highlighted. Their travel time is approximated by the piecewise linear model. We can see that Ped. 2 has lower desired velocity in free regime but higher ability to push through the crowd in comparison to Ped. 4.}
\label{fig:TT-Nmean}
\end{figure}

One of the investigated quantities is the travel-time $TT=T_\mathrm{out}-T_\mathrm{in}$ denoting the length of the interval a pedestrian spent in the room between the entrance at $T_\mathrm{in}$ and the egress at $T_\mathrm{out}$. To capture the pedestrians behaviour under variety of conditions, the travel time is investigated with respect to the average number of pedestrians in the room $N_\mathrm{mean}$ defined as
\begin{equation}
\label{eq:Nmean}
	N_\mathrm{mean}=\frac{1}{T_\mathrm{out}-T_\mathrm{in}}\int_{T_\mathrm{in}}^{T_\mathrm{out}}N(t)\mathrm{d}t\,,
\end{equation}
where $N(t)$ stands for the number of pedestrians in the room at time $t$.
Figure~\ref{fig:TT-Nmean} shows the scatter plot of all pairs $(N_\mathrm{mean},TT)$ gathered over all runs of experiment and all participants.

The reaction of participants on the occupancy of the room significantly differs. There are two basic characteristics of that can be extracted: the mean travel time in the free-flow regime (0 - 7 pedestrians) and the slope of the travel-time dependence on the number of pedestrians in the congested regime (10 - 45 pedestrians). The former is given by the desired velocity, the latter reflects the ability to push through the crowd, referred to as the aggressiveness. This observation corresponds to the piece-wise linear model for each pedestrian
\begin{equation}
\label{eq:plm}
	TT=\frac{S}{v_0(i)} + \mathbf{1}_{\{N>7\}}(N-7) \cdot \mathrm{slope}(i) + \mathrm{noise}\,
\end{equation}
where $S=7.2$~m, $v_0(i)$ is the free-flow velocity of the pedestrian $i$, $\mathrm{slope}(i)$ is the unique coefficient of the linear model for pedestrian $i$. The breakpoint $N=7$ depends from the room geometry. The weighted mean of the $R^2$ value of the model~(\ref{eq:plm}) is 0.688.

Detailed description of the experiment and its analyses has been presented at the conference TGF 15 and will be published in the proceedings~\cite{BukHraKrbTGF15}. Videos capturing the exhibition of the aggressive behaviour are available at \url{http://gams.fjfi.cvut.cz/peds}.

\section{Model Definition}

The model adapts the principle of the known Floor-Field cellular model. The playground of the model is represented by the rectangular two-dimensional lattice $\mathbb{L}\subset\mathbb{Z}^2$ consisting of cells $x=(x_1,x_2)$. Every cell may be either occupied by one agent or empty. Agents are moving along the lattice by hopping from their current cell $x\in\mathbb{L}$ to a neighbouring cell $y\in\mathcal{N}(x)\subset\mathbb{L}$, where the neighbourhood $\mathcal{N}(x)$ is Moore neighbourhood, i.e.,
$
	\mathcal{N}(x)=\left\{y \in \mathbb{L};~\max_{j=1,2}|x_{j}-y_{j}|\leq1\right\}\,.
$

\subsection{Choice of the New Target Cell}

Agents choose their target cells $y$ from $\mathcal{N}(x)$ stochastically according to probabilistic distribution $P\left(y\mid x; \mathrm{~state~of~} \mathcal{N}(x)\right)$, which reflects the ``attractiveness'' of the cell $y$ to the agent. The ``attractiveness'' is expressed by means of the static field $S$ storing the distances of the cells to the exit cell $E=(0,0)$, which is the common target for all agents. For the purposes of this article, the euclidean distance has been used, i.e., $S(y)=\sqrt{|y_{1}|^2+|y_{2}|^2}$. Then it is considered
$P(y \mid x)\propto\exp\{-k_SS(y)\}$, for $y\in\mathcal{N}(x)$. Here $k_S\in[0,+\infty)$ denotes the parameter of sensitivity to the field $S$.

The probabilistic choice of the target cell is further influenced by the occupancy of neighbouring cells and by the diagonality of the motion. An occupied cell is considered to be less attractive, nevertheless, it is meaningful to allow the choice of an occupied cell while the principle of bonds is present (explanation of the principle of bonds follows below). Furthermore, the movement in diagonal direction is penalized in order to suppress the zig-zag motion in free flow regime and support the symmetry of the motion with respect to the lattice orientation.

Technically this is implemented as follows. Let $O_{x}(y)$ be the identifier of agents occupying the cell $y$ from the point of view of the agent sitting in cell $x$, i.e. $O_{x}(x)=0$ and for $y\neq x$ $O_{x}(y)=1$ if $y$ is occupied and $O_{x}(y)=0$ if $y$ is empty. Then $P(y\mid x)\propto(1-k_OO_x(y))$, where $k_O\in[0,1]$ is again the parameter of sensitivity to the occupancy ($k_O=1$ means that occupied cell will never be chosen). Similarly can be treated the diagonal motion defining the diagonal movement identifier as $D_{x}(y)=1$ if $(x_{1}-y_{1})\cdot(x_{2}-y_{2})\neq0$ and $D_{x}(y)=0$ otherwise. Sensitivity parameter to the diagonal movement is denoted by $k_D\in[0,1]$ ($k_D=1$ implies that diagonal direction is never chosen).

The probabilistic choice of the new target cell can be than written in the final form
\begin{equation}
\label{eq:Pxy}
	P(y\mid x)=\frac{\exp\big\{-k_{S}S(y)\big\}\big(1-k_{O}O_{x}(y)\big)\big(1-k_{D}D_{x}(y)\big)}{\sum_{z\in\mathcal{N}(x)}\exp\big\{-k_{S}S(z)\big\}\big(1-k_{O}O_{x}(z)\big)\big(1-k_{D}D_{x}(z)\big)}\,.
\end{equation}
It is worth noting that the site $x$ belongs to the neighbourhood $\mathcal{N}(x)$, therefore the equation~(\ref{eq:Pxy}) applies to $P(x\mid x)$ as well.

\subsection{Updating Scheme}
The used updating scheme combines the advantages of fully-parallel update approach, which leads to necessary conflicts, and the asynchronous clocked scheme~\cite{CorGreNew2005PhysicaD} enabling the agents to move at different rates.  

Each agent carries as his property the \emph{own period} denoted as $\tau$, which represents his desired duration between two steps, i.e., the agent desires to be updated at times $t=k\tau$, $k\in\mathbb{Z}$. Such principle enables to model different velocities of agents, but undesirably suppresses the number of conflicts between agents with different $\tau$. To prevent this, we suggest to divide the time-line into isochronous intervals of the length $h>0$. During each algorithm step $k\in\mathbb{Z}$  such agents are updated, whose desired time of the next actualization lies in the interval $\big[kh,(k+1)h\big)$. A wise choice of the interval length $h$ in dependence on the distribution of $\tau$ leads to the restoration of conflicts in the model. It worth noting that we use the concept of adaptive time-span, i.e., the time of the desired actualization is recalculated after each update of the agent, since it can be influenced by the essence of the motion, e.g., diagonal motion leads to a time-penalization, since it is $\sqrt{2}$ times longer. For more detail see e.g.~\cite{BukHraKrb2014LNCS}. This is an advantage over the probabilistic approach introduced in~\cite{BanCroViz2015TGF}.

\subsection{Principle of Bonds}

The principle of bonds is closely related to the possibility of choosing an occupied cell.  An agent who chooses an occupied cell builds a bond to the agent sitting in the chosen cell. This bond lasts until the motion of the blocking agent or until the next activation of the bonded agent. The idea is that the bonded agents attempt to enter their chosen cell immediately after it becomes empty.

\subsection{Aggressiveness and Solution of Conflicts}

The partially synchronous updating scheme of agents leads to the kind of conflicts that two ore more agents are trying to enter the same cell. This occurs when more agents choose as their target cell the same cell, or when more agents are bonded to the same cell, which becomes empty. The mechanism of the conflict solution is the same in both cases. Each agent carries an information about his ability to ``win'' conflicts which is here referred to as aggressiveness and denoted by letter $\ag\in[0,1]$. The conflict is always won by agents with highest $\ag$.

\begin{figure}
\centering
	\includegraphics[scale=1]{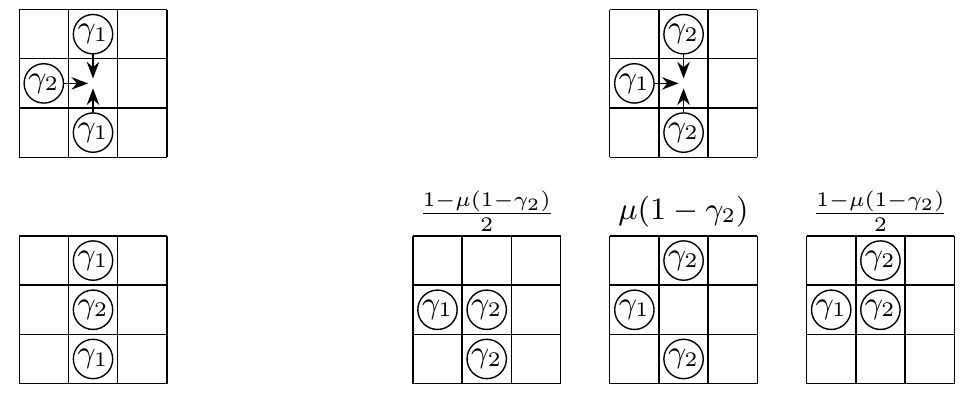}
\caption{Conflict solution for $\ag_1<\ag_2$. Left: More aggressive wins the conflict over two less aggressive. Right: The conflict of two more aggressive can resolve by the blocking of the movement.}
\label{fig:conflict}
\end{figure}

If there are two or more agents with the highest $\ag$, the friction parameter $\mu$ plays a role. In this article we assume that the higher is the aggressiveness $\ag$, the less should be the probability that none of the agents wins the conflict. Therefore, the conflict is not solved with probability $\mu(1-\ag)$ (none of the agents move). With complement probability $1-\mu(1-\ag)$ the conflict resolves to the motion of one of the agents. This agent is chosen randomly with equal probability from all agents with involved in the conflict having the highest $\ag$. The mechanism of the friction can be easily modified. An example of conflict solution is depicted in Figure~\ref{fig:conflict}.

\section{Impact of the Aggressiveness Element}

The effect of the aggressiveness has been studied by means of the simulation. Results stressed in this article come from the simulations with parameters given by Table~\ref{tab:param}. The values of $\tau$ and $\ag$ are distributed among agents uniformly and independently on each other.

\begin{table}
\caption{Values of parameters used for simulation.}
\label{tab:param}
\centering
	\begin{tabular}{p{1.2cm}p{1.2cm}p{1.2cm}p{1.2cm}p{1.2cm}p{1.2cm}p{1.2cm}}
	\hline
		$k_S$ \centering & \centering $k_O$ & \centering $k_D$ &\centering $h$ & \centering $\mu$ & \centering $\tau$ & ~~~~$\ag$\\
		3.5 \centering &\centering 1 & \centering 0.7 & \centering 0.1~s &\centering 0.5 & \centering $\left\{.25, .4\right\}$ &  ~~$\{0,1\}$\\
		\hline
	\end{tabular}
\end{table}

The simulation set-up has been designed according to the experiment, i.e., the room of the size $7.2~\mathrm{m} \times 4.4~\mathrm{m}$ has been modelled by the rectangular lattice 18 sites long and 11 sites wide. The size of one cell therefore corresponds to $0.4~\mathrm{m}\times 0.4~\mathrm{m}$. The exit is placed in the middle of the shorter wall, the open boundary is modelled by a multiple entrance on the opposite wall. New agents are entering the lattice stochastically with the mean inflow rate $\alpha$~[pedestrians/second]. The inflow rate is a controlled parameter. For more detailed description of the simulation we refer the reader to~\cite{BukHra2014LNCS}.

It has been shown that such system evinces the boundary induced phase transition from the free flow (low number of agents in the lattice) to the congestion regime (high number of pedestrians in the lattice) via the transient phase (number of pedestrians fluctuating between the low and high value). Therefore, wise choice of different inflow rates $\alpha$ covering the all three phases, enables us to study the dependence of the travel time $TT$ on the average number of agents in the lattice $N_\mathrm{mean}$. When simulating with parameters from Table~\ref{tab:param}, the correct choice of inflow rate is $\alpha\in[1,3]$.

\begin{figure}[htb]
	\includegraphics[width=\textwidth]{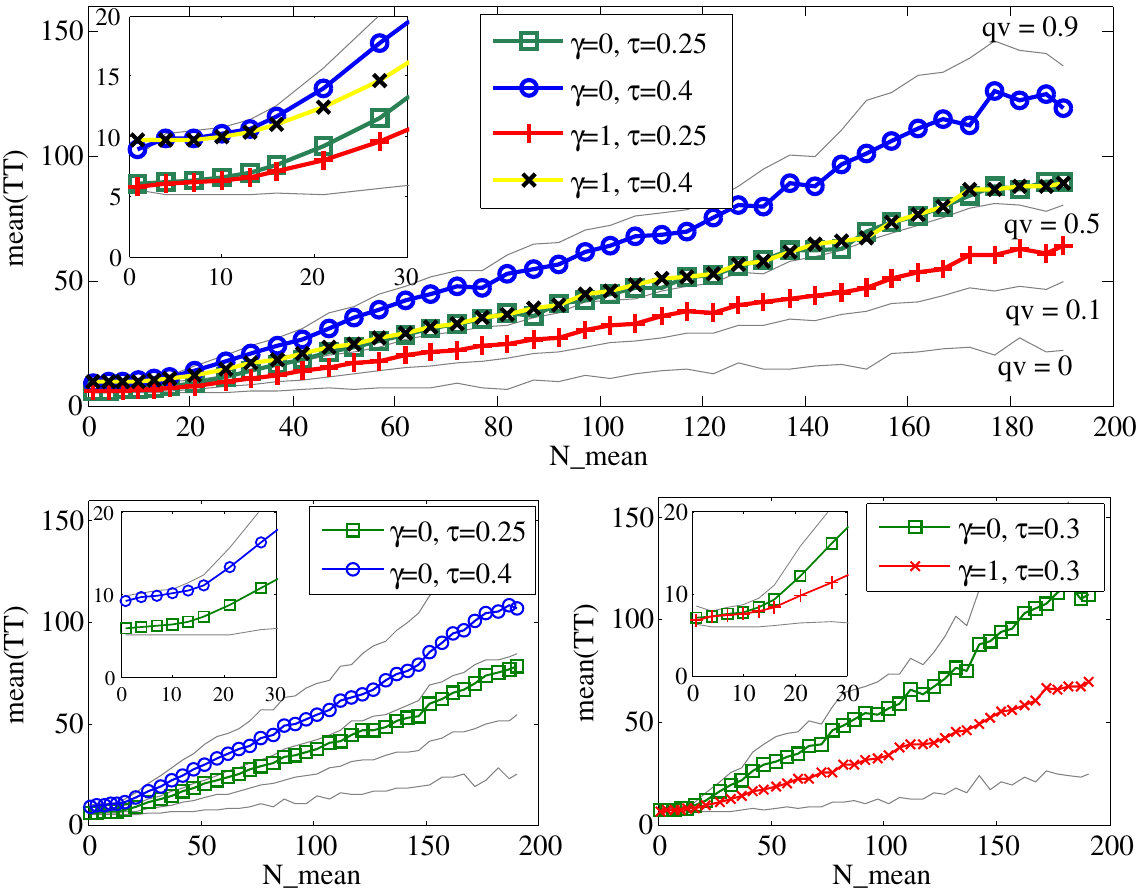}
\caption{Dependence of the mean travel time $TT$ on the average occupancy $N_\mathrm{mean}$ for each group of agents. Gray lines represent the quantiles of the travel time regardless to the groups. Top: heterogeneity in both, $\gamma$ and $\tau$. Bottom left: heterogeneity in $\tau$. Bottom right: heterogeneity in $\gamma$.}
\label{fig:TT-N}
\end{figure}

Figure~\ref{fig:TT-N} shows the dependence of the travel time $TT=T_\mathrm{out}-T_\mathrm{in}$ on the average number of agents in the lattice $N_\mathrm{mean}$ calculated according to~(\ref{eq:Nmean}). Measured data consisting of pairs $(N_\mathrm{mean},TT)$ are aggregated over simulations for inflow rate values $\alpha\in\{1, 1.5, 1.8, 2.0, 2.3, 2.7, 3.0\}$; for each inflow $\alpha$ twenty runs of the simulation have been performed. Each run simulates 1000~s of the introduced scenario starting with empty room. Agents were distributed into four groups according their own period $\tau$ and aggressiveness $\ag$, namely ``fast aggressive'' ($\tau=0.25$, $\ag=1$), ``fast calm'' ($\tau=0.25$, $\ag=0$), ``slow aggressive'' ($\tau=0.4$, $\ag=1$), and ``slow calm'' ($\tau=0.4$, $\ag=0$).

In the graph of the Figure~\ref{fig:TT-N} we can see the average travel time for each group calculated with respect to the occupancy of the room. It is evident that for low occupancy up to 10 agents in the room the mean travel time for each group levels at a value corresponding to the free flow velocity given by the own updating period. For the occupancy above 20 agents in the lattice, the linear growth of the mean travel time with respect to $N_\mathrm{mean}$ is obvious. Furthermore, the average travel time for fast-calm corresponds to the travel time of slow aggressive. The Figure~\ref{fig:TT-N} shows two auxiliary graphs presenting the dependence of $TT$ on $N_\mathrm{mean}$ for systems with homogeneity in $\gamma$ (left) or in $\tau$ (right). From the graphs we can conclude that the heterogeneity in aggressiveness $\gamma$ reproduces the desired variance in the slope of the graph without the non-realistic high variance in free flow generated by the heterogeneity of own updating frequency.

\begin{figure}
	\includegraphics[width=\textwidth]{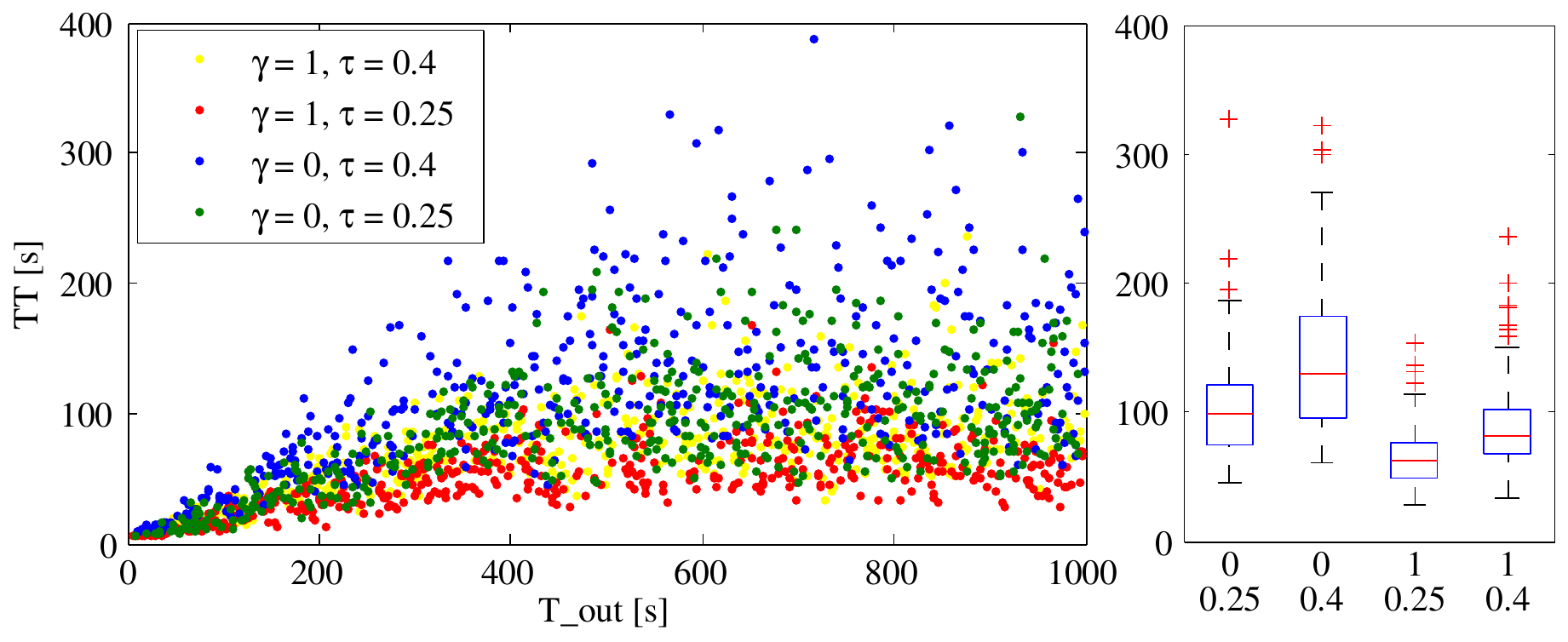}
\caption{Left: Development of travel time $TT$ in time for one run of the simulation. The value $TT$ is plotted against the time of the exit $T_\mathrm{out}$ to ensure that values corresponding to the same time stem from similar conditions near the exit. Inflow rate $\alpha=3$~ped/s. The agent group is indicated by the color. Right: Box-plots of the travel time for entries with $T_\mathrm{in}>500$~s (i.e. in the steady state).}
\label{fig:TT-Tout}
\end{figure}

The influence is even more evident from the graph in Figure~\ref{fig:TT-Tout} representing a plot of all  travel time entries with respect to the time of the exiting $T_\mathrm{out}$. Right graph shows the box-plots of the travel time for four groups measured after 500~s from the initiation, i.e., in the steady phase of the system. We can see that in this view, the aggressiveness plays more important role that the desired velocity of agents.

\section{Conclusions and Future Work}

The article introduced a parameter of aggressiveness as an additional characteristics of agents in the Floor-Field model with adaptive time. This parameter is understood as an ability to win conflicts. Therefore the heterogeneity of agents is given by their desired velocity (determined by the own period $\tau$) and their ability to win conflicts referred to as the aggressiveness.

The simulation study shows that the aggressiveness has significant influence in the regime with high occupation of the room, i.e., in the dense crowd, and on the other hand has no effect in the free flow, as desired. The linear dependence of the travel time on the number of pedestrians in the agents neighbourhood seems to be a good tool how to measure the ability of agents/pedestrians to push through the crowd. The independence of this ability on the desired velocity of agents is very important to mimic the aspect that some ``fast'' pedestrians can be significantly slowed down by the crowd while some ``slow'' pedestrians can push through the crowd more effectively.

We believe that such feature can be very useful in the simulation of the evacuation or egress of large complexes as e.g. football stadiums, where the less aggressive pedestrians (parents with children, fragile women) can be slowed down and leave the facility significantly later than the average. The model reproduces this aspect even in the case of the homogeneity in own period $\tau$.

In the future we plan to study this aspect in more detail. Mainly we would like to focus on the joint distribution of the desired velocity $\tau$ and the aggressiveness $\ag$ among the population and study its impact by means of the proposed model.

\subsection*{Acknowledgements.} This work was supported by the Czech Science Foundation under grants GA13-13502S (P. Hrab\'ak) and GA15-15049S (M. Buk\'a\v cek). Further support wa provided by the CTU grant SGS15/214/OHK4/3T/14.

\bibliographystyle{splncs03}
\bibliography{CP92_references}

\end{document}